\documentclass[twocolumn,prb,amsmath,amssymb,showpacs,superscriptaddress,floatfix]{revtex4-1}

\usepackage[hypertex]{hyperref}
\usepackage{graphicx}
\usepackage{srctex}
\DeclareMathOperator{\Tr}{Tr}  
\begin{document}

\title{Full \textit{versus} first stage replica symmetry breaking in spin glasses.
}
\author{T.I. Schelkacheva}
\affiliation{Institute for High Pressure Physics, Russian Academy of Sciences, Troitsk 142190, Moscow Region, Russia}
\author{E.E. Tareyeva}
\affiliation{Institute for High Pressure Physics, Russian Academy of Sciences, Troitsk 142190, Moscow Region, Russia}
\author{N.M. Chtchelkatchev }
\affiliation{Institute for High Pressure Physics, Russian Academy of Sciences, Troitsk 142190, Moscow Region, Russia}
\affiliation{Materials Science Division, Argonne National Laboratory, Argonne, Illinois 60439, USA}

\date{\today}

\begin{abstract}
A short survey  is presented on spin--glass--like states characteristics in complex nonmagnetic systems. We discuss the interplay of the interaction structure and symmetry with the classification scenarios of the replica symmetry breaking. It is shown that the kind of the transition to the nonergodic state depends not only on the presence or absence of the reflection symmetry but on the number of interacting operators and their individual characteristics.
\end{abstract}

\maketitle

\section{Introduction\label{Sec:Intro}}

The reflection symmetry plays the crucial role defining the character of phase transition in nonrandom mean--field (MF) models.\cite{LLSt} Generally speaking the presence of the terms without reflection symmetry usually results in the first order phase transition, while in the absence of such terms the transition is of the second order. In the case of random MF systems the absence of reflection symmetry also leads to a special form of the free energy functional that differs from the symmetrical case. As a consequence, the scenarios of replica symmetry breaking (RSB) are different for these two cases. However, not only symmetry determines the transition to
nonergodic state. Extending the class of models permits considering the role of different factors in the scenarios of appearance of SG-type states. In this paper we try to use our recent results for different models for spin--glass--like states in complex nonmagnetic systems (see Ref.[\onlinecite{TMFNNB}] for a review) to investigate how the interaction type correlates with the spin glass (SG) behavior.


The theory of spin glasses developed as an attempt to describe unordered equilibrium freezing of spins in actual dilute magnetic systems with disorder and frustration. This problem was soon solved at the mean-field level Edwards and Anderson,\cite{EA} Sherrington and Kirkpatrick,\cite{sk} Almeida and Thouless,\cite{A-T} and Parisi~\cite{par1,par2} [see Ref.~\onlinecite{book} for a review]. The Sherrington--Kirkpatrick approach to the spin-glass theory starts from the Hamiltonian
\begin{equation}\label{SK}
H=-\frac{1}{2}\sum_{i\neq j}J_{ij} U_i U_j.
\end{equation}
It describes Ising spins $U$ located on the lattice sites $i$. The quenched interactions $J_{ij}$ are distributed with the Gaussian probability,
\begin{equation} \label{Gauss}
P(J_{ij})=\frac{1}{\sqrt{2\pi J}}\exp\left[-\frac{(J_{ij}-J_0)^{2}}{2J^{2}}\right],
\end{equation}
where $ J=\tilde{J}/\sqrt{N}$ , $J_{0}=\tilde{J_0}/N$ and $N$ s the number of sites. To perform averaging over disorder in this case one has to average the quenched free energy $F$ rather than the partition sum $Z$ itself. Such averaging is usually performed using the replica method. After that the free energy becomes the function of the order parameters that depend on replica indices:
\begin{gather}
F = F(x^{\alpha }, q^{\alpha \beta }),
\\
x^{\alpha }=\frac{1}{N}\sum\limits_{i=1}^{N}U_i^{\alpha },\qquad
q^{\alpha \beta }=\frac{1}{N}\sum\limits_{i=1}^{N}U_i^{\alpha }U_i^{\beta }.
\end{gather}
The free energy $F(x^{\alpha }, q^{\alpha \beta })$ has an extremum for the replica symmetric (RS) solution when all $q^{\alpha \beta }$ are equal. However this state is unstable under RSB. Parisi proposed the method of RSB step by step with the limit full RSB (FRSB) when  $q^{\alpha\beta }$ becomes a continuous function of a parameter $x$. This approach allows to describe the main features of the experiments on spin glasses. Namely, in the framework of the equilibrium approach, the spin glass phase with  qualitatively correct boundaries was obtained and the difference in the behavior of magnetic susceptibility in field--cooled and zero--field--cooled cases was explained.

So, the problem of theoretical description of SG {\it per se} was solved in principle and during that time different other models appeared without any connection to real experiments and real physical systems. The main feature of these models was the absence of time reversal symmetry  -- in contrast to the SK model. The most investigated model among those are the $p$-spin models and the Potts models, considered, for example, in Refs.~\onlinecite{goeld,kirkp,rev3,Gardner}. The spherical $p$-spin model [see, e.g., Ref.~\onlinecite{CrSo92}] was believed for a long time to be a generic for this class of models. From the point of view of RSB the main feature of this model is the stability of the first step of RSB (1RSB) down to zero temperature. Also, the order parameter behaves stepwise. Although this model was not aimed to describe any actual glass it appears to be very interesting because its behavior gives a scenario for real liquid-glass transition: two critical temperatures, the number of metastable states similar to that obtained in numerical modeling. It should be noted that the structure of the dynamical equations for the correlation functions are identical for the supercooled liquids in the mode-coupling theory and for the $p$-spin spherical SG model.\cite{kirkp}

Based mainly on the investigations of these two models -- SK and $p$-spin spherical -- a conclusion appears in the literature attributing the two classes of universality to the models with and without reflection  symmetry.
In the disordered case a number of attempts were made to formulate a kind of universality rules based on the mean-field investigation of the model systems with the random interactions.\cite{kirkp,goeld}

Looking now in general at the free energy series over the glass order parameter we see that the series contain explicitly the terms which can be classified by the reflection symmetry,
\begin{multline}
\frac {\Delta F^s}{NkT}=\lim_{n \rightarrow 0}\frac{1}{n} \sum \left[... +a_3 \delta q^{\alpha\beta} \delta q^{\beta\gamma } \delta q^{\gamma \alpha}+\right.
\\\left.
a'_4  (\delta q^{\alpha\beta})^4 +a_4 \delta q^{\alpha\beta} \delta q^{\beta\gamma } \delta q^{\gamma \delta } \delta q^{\delta \alpha }...\right],
\end{multline}
and the part without the reflection symmetry: the terms with three identical replica indices:
\begin{multline}
\frac {\Delta F^{ns}}{NkT}= \lim_{n \rightarrow 0}\frac{1}{n} \sum\left[...+b_3 (\delta q^{\alpha\beta})^3+... +\right.
\\\left.
b_4 \delta q^{\alpha\beta} \delta q^{\beta\gamma } \delta q^{\gamma \alpha} \delta q^{\delta \alpha }...\right].
\end{multline}
Thus, a natural question arises: whether there can be made a general statement regarding the behavior of SG models with and without reflection  symmetry? And do all models of the first type behave in fact as SK model  and all models of the second type as $p$-spin spherical model? In this  paper we try to answer this question. We prove that  for arbitrary models with the reflection symmetry the Parisi FRSB always takes place. In the absence of the reflection symmetry the situation is not so definite and the behavior of the system depends on some additional characteristics. In any case it is not always similar to that of he $p$-spin spherical model, as it was usually believed:  we give the counterexamples.

\section{Generalized SK model: FRSB\label{Sec:Full}}

\subsection{SK-model with the reflection symmetry}
In this case it occurs to be possible to prove a kind of a theorem.

First, we consider a generalized model with reflection symmetry with the  
Hamiltonian (\ref{SK}), with the interactions distributed according to Eq.(\ref{Gauss}) and with the arbitrary diagonal operators  $U$. The reflection symmetry implies that for any integer $k$,
\begin{equation}
\Tr\left[{U}^{(2k+1)}\right]=0. \label{odd}
\end{equation}

The saddle point conditions for the free energy averaged over disorder produces the glass order parameter
\begin{equation}\label{qdef}
q^{\alpha\beta}={\Tr\left[U^{\alpha}U^{\beta} \exp\left(\theta\right)\right]}/{\Tr\left[\exp\left(\theta\right)\right]},
\end{equation}
and the auxiliary order parameter
\begin{equation}\label{wdef}
 w^{\alpha}={\Tr\left[(U^{\alpha})^2 \exp\left(\theta\right)\right]}/{\Tr\left[\exp\left(\theta\right)\right]}.
\end{equation}
Here
\begin{equation}\label{six}
\theta=\frac{t^2}{2}  \sum_{\alpha}w^{\alpha}(U^{\alpha})^2+t^2 \sum_{\alpha>\beta}q^{\alpha\beta}U^{\alpha}U^{\beta},
 \end{equation}
where $t=\tilde{J}/kT$ and we choose $J_0=0$ for simplicity.

In the RS-approximation we find the (trivial) solution $q_{\mathrm{RS}} = 0$. The bifurcation condition looks like in this case:
\begin{equation} \label{ten}
1-t_c^2 w^2(t_c) = 0.
\end{equation}
This equation coincides with $\lambda_{\mathrm{repl}(\mathrm{RS})} = 0$ [see, e.g., Ref.~\onlinecite{book}]. It is very important that it is zero solution that bifurcates.

Investigating $1\mathrm{RSB}$, $2\mathrm{RSB}$, $3\mathrm{RSB}$, and so on, we see that the equations for the glass order parameters always contain the quantity
\begin{gather}
{\Tr[U \exp (\theta_{n\mathrm{RSB}})]}/{\Tr[\exp (\theta_{n\mathrm{RSB}})]}.
\end{gather}
Therefore, one of the solutions of these equation is trivial at each of the $\mathrm{RSB}$-steps, and the appearance of the $n\mathrm{RSB}$ solution can be regarded as the bifurcation of the trivial $(n-1)\mathrm{RSB}$  solution. In this case, the equation  $\lambda _{n\mathrm{RSB}} = 0$ coincides with the corresponding branching condition (\ref{ten}). This means that in any case, the $n\mathrm{RSB}$ solutions at different stages of the symmetry breaking can exist at the temperature $T< T_c$ determined by this bifurcation condition, and so we always can look for $\mathrm{FRSB}$ solution. Writing the free energy as a series over $\delta q^{\alpha\beta }$ near $T_c$ (up to the fourth order of magnitude inclusively) we obtain $q(x)=cx$ in the leading approximation [similar procedure in details was described in \onlinecite{Full}]. It is also possible to write the free energy in the form of Parisi with the only difference in the boundary conditions for the Parisi function $\phi$ that now reads:
\begin{gather}
\phi (1,y) = \ln \Tr \left\{\exp \left[ ty  U + \frac {t^2}{2}\left(w-q(1)\right) U^2\right ] \right  \}.
\end{gather}
Thus we have shown that in the case of systems with reversal symmetry, the infinite $\mathrm{FRSB}$ occurs at the very point at which the $\mathrm{RS}$ solution becomes unstable. In particular, our result means that
magnetic systems of arbitrary spin with the interaction between the $z$-components behave in the same way.

\subsection{SK-model without the reflection symmetry}
We consider below the models without the reflection symmetry. [These models correspond to some real physical systems.] It implies that we have the Hamiltonian (\ref{SK})-(\ref{Gauss}) but without the condition (\ref{odd}) for the operators $U$.  It is easy to trace how the proof given above fails using the model proposed in Ref.~\onlinecite{4avtora}.

The difference between two cases is already manifested  in the $\mathrm{RS}$ approximation. In the case when the condition (\ref{odd}) is not fulfilled for the Hamiltonian (\ref{SK}) there is no trivial solution for the order parameters. The disorder smears out the first-order phase transition; hence, instead of a transition, there is a smooth increase in the order parameters (both glass and regular) as the temperature decreases. This situation is seen in experiments on orientational glass phase in $ortho-para$--hydrogen mixed crystals and in $Ar-N_2$.\cite{Silivan} These substances present mixtures of spherically symmetric molecules and momentum bearing molecules. The corresponding glass was investigated  on the base of the Hamiltonian~(\ref{SK}) with $U=Q$, where $Q=3 J^2_{z} - 2$,  ${\bf J}=1$.\cite{Luchinskaya} The RSB solution branches continuously and smoothly on cooling breaking the RS results in a transition to the nonergodic phase of the quadrupolar glass.

Another example of a SG--like phase in the molecular crystal is  the pure $para-H_2$ (or $ortho-D_2$) under pressure.\cite{ Goncharov} The  possibility of the orientational order in the systems with the initially spherically symmetric molecule states is due to the involving of higher order orbital  moments $J = 2, 4...$ under pressure. With increase of the density the anisotropic interaction potential and the crystal field grow rapidly and the energy of the many body system can be lowered taking the advantage of the anisotropic interactions. The long range orientational order appears abruptly at a fixed value of pressure through the first order phase transition just as it takes place in ortho-para mixtures when the concentration of moment bearing molecules achieves certain fixed value. In the intermediate concentration range the frustration and disorder motivate the investigation of the quadrupole glass with $J=2$. Such a theory was constructed in~\onlinecite{paraH}. The essential feature of the obtained intermediate phase is the coexistence of the orientational glass phase with the long range orientational order as it is seen in the experiment.

We consider two other models describing SG--like states in real complex nonmagnetic systems, namely, in the cluster systems. Although they are not the mixtures of the different kinds of particles with different interactions, one can find frustration and  disorder, that is the background to consider these systems in the spirit of SG theory. Now the operator $U$ in (\ref{SK}) is  replaced with continuous functions of the angles.

In Ref.~\onlinecite{C60} a model for the low-temperature transition to the orientational glass state in solid molecular $C_{60}$ was developed.  Although the molecules have nearly spherical shape, at low temperature there are two pronounced minima in the anisotropic part of the intermolecular interaction energy. It is possible to trace an analogy with the mixtures studying the role of the different types of the mutual molecular orientations. As a result, a model is constructed where the role of spin is played by certain combinations of the cubic harmonics.  The results agree well with the experimental data: the coexistence of the glass state and the long-range orientational order and the existence of a wide maximum on the curve for the orientational part of the  heat capacity. Moreover, the above model permits considering the pressure dependence of the orientational transitions.\cite{Moret}

The other model we would like to mention is the SG--like freezing of clusters of different symmetries in supercooled liquids that gives a possible description of liquid--glass transition. In Ref.\onlinecite{cluster} we use the microscopic approach based on the equations for the distribution functions which in spirit of Bogoliubov hierarchy give us an opportunity to analyze the intercluster interaction. We show that there exists a region of densities and temperatures where this interaction changes sign as a function of the cluster radius and there is hence frustration in the system. This is the base to write a Hamiltonian of the form (\ref{SK}) with different point group harmonics for $U$ and use standard methods of SG theory to describe the  real glasses.

So, we have considered a set of models with two-particle interaction where the absence of the  reflection symmetry is caused by the characteristics of the operators $U$ themselves. In this case the $\mathrm{RSB}$-solution bifurcates from the $\mathrm{RS}$ -solution smoothly, without a jump, and the coexistence of the glass order with the long range regular order takes place.

\section{Generalized $p$-spin model\label{Sec:pspin}}

We consider now a generalization of the well-known $p$-spin model \cite{Gardner} of Ising spins where spins are replaced by arbitrary diagonal operators. Then the
Hamiltonian is
\begin{equation}
H=-\sum_{{i_{1}}\leq{i_{2}}...\leq{i_{p}}}J_{i_{1}...i_{p}}
U_{i_{1}}U_{i_{2}}...U_{i_{p}}, \label{oneone}
\end{equation}
where $i=1,2,...N$, $p$ is the number of interacting particles  and $U$ is an arbitrary diagonal operator such that $\Tr U = 0$. We do not specify its form here, in order to use general expressions further. The independent  interactions have the Gaussian distribution
\begin{equation}\label{twotwo}
P(J_{i_{1}...i_{p}})=\frac{\sqrt{N^{(p-1)}}}{\sqrt{p!\pi} \widetilde{J}}\exp\left[-\frac{(J_{i_{1}...i_{p}})^{2}N^{(p-1)}}{p!{\widetilde{J}}^{2}}\right].
\end{equation}

Using the standard procedure of replica approach we obtain the free energy and the equations for the order parameters (\ref{qdef}) and (\ref{wdef}), but now with,
\begin{multline}\label{six}
\theta=p\frac{t^2}{2} \sum_{\alpha>\beta}(q^{\alpha\beta})^{(p-1)}U^{\alpha}U^{\beta}+
\\
p\frac{t^2}{4}\sum_{\alpha}{(w^{\alpha})}^{(p-1)}(U^{\alpha})^2.
 \end{multline}

We perform the first stage $\mathrm{RSB}$  [$n$ replicas are divided into $n/m_1$ groups each containing $m_1$ replicas] and obtain the free energy in the form
\begin{multline}\label{1rsbf}
F_{1\mathrm{RSB}}=-NkT[m_{1} t^2(p-1)\frac{r_{1}^p}{4}
\\
+ (1-m_{1})(p-1) t^2\frac{(r_{1}+v_{1})^p}{4}-t^2(p-1)\frac{w_{1}^p}{4}
\\
+ \frac{1}{m_{1}}\int dz^G\ln \int ds^G \left[\Tr\exp{\theta_{1\mathrm{RSB}}}\right]^{m_{1}}].
\end{multline}
Here $q^{\alpha \beta }= r_1$ if $\alpha $ and $\beta $ are from the different groups and $q^{\alpha \beta}= r_{1}+v_{1}$ otherwise,
\begin{multline}\label{teta1}
\theta_{1\mathrm{RSB}}=zt\sqrt{\frac{p{r_{1}}^{(p-1)}}{2}}\,U
\\
+st\sqrt{\frac{p[{(r_{1}+v_{1})}^{(-1)}-{r_{1}}^{(p-1)}]}{2}}\,U
\\
+t^2\frac{p[w_{1}^{(p-1)}-{(r_{1}+v_{1})}^{(p-1)}]}{4}U^2.
\end{multline}

\begin{figure}[t]
\begin{center}\includegraphics[width=0.485\textwidth]{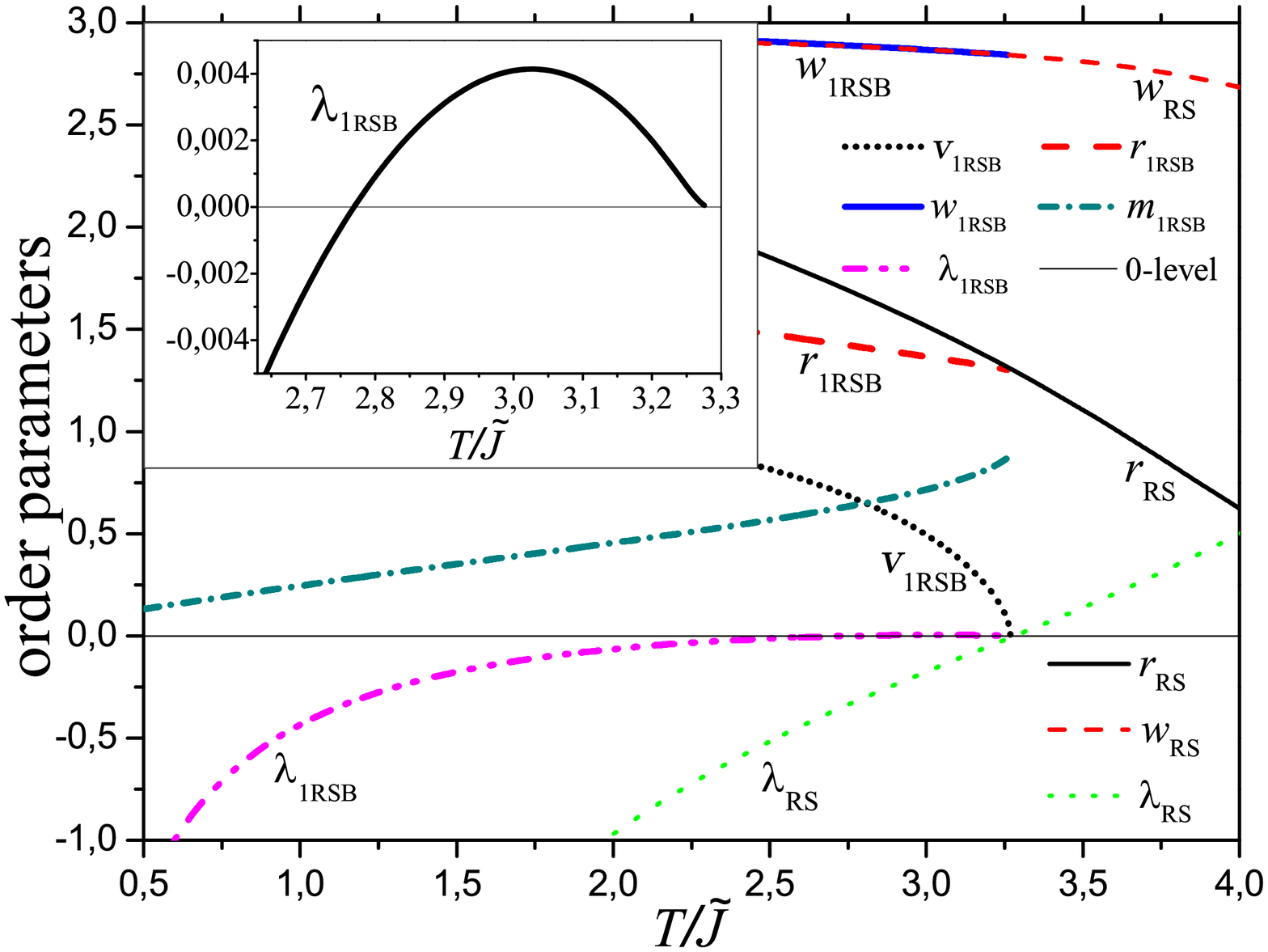}\end{center}
\caption{\label{fig:fig1} (Color online) Temperature dependence of the glass order parameters for the quadrupole glass with three particle interaction for $J=1$. The $\mathrm{RSB}$ occurs at the temperature corresponding to the condition $\lambda_{(\mathrm{RS})\mathrm{repl}}=0$.}
\end{figure}

\begin{figure}[t]
\includegraphics[width=0.48\textwidth]{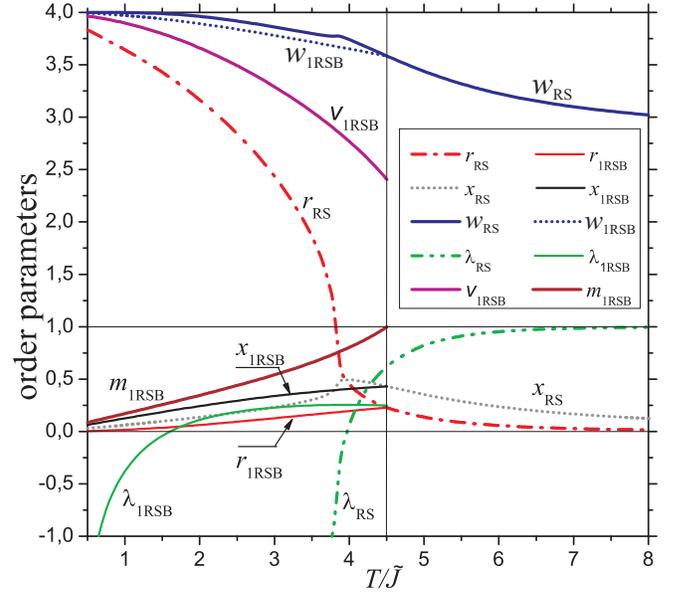}%
\caption{\label{fig:fig2}(Color online)
 Temperature dependence of the order parameters for the quadrupole glass with three particle interaction for $J=2$. The transition $\mathrm{RS}-1\mathrm{RSB}$ takes place at the point defined by the condition $m=1$. The glass order parameter $v_{1}$ has the jump at this point.Here $x$ is the regular orientational order parameter.}
\end{figure}

We performed the detailed calculations for two models with $p=3$ and where $U$ were represented by the quadrupolar moment with $J=1,2$. The results of the calculations are illustrated in Fig.~\ref{fig:fig1} ($J=1$) and Fig.~\ref{fig:fig2} ($J=2$). The stability of the $1\mathrm{RSB}$ solution against further $\mathrm{RSB}$ was checked in the standard way\cite{TMFNNB,paraH} looking at the positive values of $\lambda_{(1\mathrm{RSB})\mathrm{repl}}$ [defined as the bifurcation point where the non-zero order parameter $v_2$ in $2\mathrm{RSB}$ appears]:
\begin{gather}\begin{split}\label{lambda}
\lambda_{(1\mathrm{RSB})\mathrm{repl}}=1-\frac{t^2}{2} p(p-1)(r_{1}+v_{1})^{(p-2)}&\times
\\
\int dz^G\int ds^G[\Tr\exp{\theta _{1\mathrm{RSB}}}]^{m_{1}} &\times
\\
\left\{\frac{\Tr(U^2 \exp{\theta _{1\mathrm{RSB}}})} {\Tr\exp{\theta_{1\mathrm{RSB}}}}-\left[\frac{\Tr(U \exp{\theta_{1\mathrm{RSB}}})} {\Tr\exp{\theta_{1\mathrm{RSB}}}}\right]^2\right\}^2 &\times
\\
\left\{\int ds^G (\Tr\exp{\theta_{1\mathrm{RSB}}})^{m_{1}}\right\}^{-1}&.
\end{split}\end{gather}

So, in the case of the three--particle interaction between quadrupoles with $J=2$ as well as with $J=1$, the first stage $\mathrm{RSB}$ is stable only in the finite region of temperatures and not down to zero temperature [as was supposed in Ref.~\onlinecite{Gardner}]. This is the key result. 
As concerns the models with the multiple interactions, this property was investigated for the Potts model with three states in Ref.~\onlinecite{grpr} [see, also Ref.~\onlinecite{JaKl10}] and for more complicated models with two interactions in Refs.~\onlinecite{crile1,crile2,crile3,crile4}.

\section{Conclusion\label{Sec:sum}}

We considered the behavior of complex spin-glass-like systems. The set of physical systems can be divided into two classes depending on whether the reflection symmetry is present or not.

We have shown that in the systems with the reflection symmetry the infinite $\mathrm{FRSB}$ takes place at the very point where the $\mathrm{RS}$ solution becomes unstable. This behavior is well known for the SK spin model. In particular, our result means that magnetic systems of arbitrary spin with the interaction between the $z$ spin-components behave in the same way.

If there is no reflection symmetry then the situation is not so definite. The behavior of the system depends on additional characteristics. An important property of such systems is the absence of a trivial RS-solution. We have considered a set of models with the two-particle interaction where the absence of the reflection symmetry is related to the structure of the $U$-operators. In this case the $\mathrm{RSB}$ solution bifurcates from the $\mathrm{RS}$ solution smoothly, without a jump. The jump appears in the 3-quadrupole glass model with $J=2$. The coexistence of the glass order with the long range regular order takes place in all the cases.

The properties of the models considered in our paper are not similar to the properties of the $p$-spin spherical model as follows from three counterexamples. We have shown that in these cases under certain additional conditions there exists a finite domain of stability for the $1\mathrm{RSB}$ order parameters. This was apparently first shown for simple nonspherical models in Refs.\onlinecite{TMFNNB,paraH} This effect was discovered for the Potts model with three states in Ref.~\onlinecite{grpr} earlier. The $\mathrm{FRSB}$  is attained as a result of several successive transitions taking place as the temperature decreases.

\begin{acknowledgments}
The authors thank V.N. Ryzhov and N. Gribova for useful discussions. This work was supported in part by the Russian Foundation for Basic Research (08-02-00781), the Dynasty Foundation (N.C.), the Federal programs and the Program of the Presidium of the Russian Academy of Sciences (E.T., T.S. and N.C.), and the
U.S. Department of Energy Office of Science under the Contract No. DE-AC02-06CH11357.
\end{acknowledgments}

\end{document}